\documentclass[12pt,a4paper]{article}

\usepackage[T1]{fontenc}
\usepackage[utf8x]{inputenc}

\usepackage{lmodern} 
\usepackage[top=1in,left=1in,footskip=1in]{geometry}

\usepackage{graphicx} 
\usepackage{authblk}

\usepackage[aboveskip=1pt,labelfont=bf,labelsep=period,justification=raggedright,singlelinecheck=off]{caption}

\makeatletter
\renewcommand{\@biblabel}[1]{\quad#1.}
\makeatother

\date{}

\usepackage{lastpage,fancyhdr,graphicx}
\usepackage{epstopdf}
\usepackage{amsmath,amssymb}

\usepackage{setspace}
\usepackage{authblk} 
\begin{document}



\title{{\bf Speed-detachment tradeoff and its effect on track bound transport of single motor protein}}
\author{Mohd Suhail Rizvi}
\affil{\small Department of Biological Sciences and Bioengineering,\\ Indian Institute of Technology Kanpur, Kanpur 208016, India.}
\date{}
\clearpage
\maketitle
\thispagestyle{empty} 
Current Address: Laboratoire Interdisciplinaire de Physique, \par 140 Rue de la Physique, 38402 Grenoble, France. \par
Corresponding Author: Mohd Suhail Rizvi\par
Email: mohd-suhail.rizvi@univ-grenoble-alpes.fr\par
Phone: +33 783956336\par
{\bf Short title:} Spontaneous motor protein detachment from track \par




\newpage
\section*{Abstract}
The transportation of the cargoes in biological cells is primarily driven by the motor proteins on filamentous protein tracks. The stochastic nature of the motion of motor protein often leads to its spontaneous detachment from the track. Using the available experimental data, we demonstrate a tradeoff between the speed of the motor and its rate of spontaneous detachment from the track. Further, it is also shown that this speed-detachment relation follows a power law where its  exponent dictates the nature of the motor protein processivity. We utilize this information to study the motion of motor protein on track using a random-walk model. We obtain the average distance travelled in fixed duration and average time required for covering a given distance by the motor protein. These analyses reveal non-monotonic dependence of the motor protein speed on its transport and, therefore, optimal motor speeds can be identified for the time and distance controlled conditions.

{\bf Keywords:} motor protein, processivity, detachment

{\bf Acknowledgements:} I would like to thank Dr. Sovan Das for his support and Sunit K. Gupta, Aditi Nag, Manu Vajpai and Mishtu Mukherjee for their valuable comments. 


\maketitle

\newpage
\section{\label{sec:intro}Introduction}
Motor proteins are deployed for the directional intra-cellular transportation. These proteins move along the filamentous tracks, formed of actin and tubulin proteins, at the expense of the energy released from ATP hydrolysis \cite{lodish,stryer_biochem}. The prominent examples of these motor proteins include kinesin, dynein and myosin which move along the microtubules and actin filaments. Kinesin and dynein motors move in the opposite directions on microtubules, whereas myosins travel towards the positive end of the F-actin \cite{lodish,stryer_biochem,pmid9335494}. 
The dynamics of the motor protein transport along the tracks has been studied extensively, and the aspect of the spontaneous motor protein detachment, that is without application of any external force, from the tracks has also been under investigation \cite{pmid12208993,pmid10753125,pmid19278641} (and references therein). The transport of the track bound motor protein and their detachment from it has been experimentally investigated by tracking of single motors in physiological \cite{pmid16873064,pmid10024239,pmid19450497,pmid11707568,pmid19617538,pmid10557288,pmid22231401,pmid19823565} and non-physiological conditions \cite{nara_2006,pmid10860848} as well as of their engineered counterparts \cite{pmid25109325,pmid18805095,pmid11086010}. Similarly, modeling approaches have also studied the influence of the spontaneous motor detachment \cite{pmid23730145,pmid16486411,pmid17142285,pmid21230234} and binding \cite{pmid8785281} with the track on its overall transport properties. In all of these works the distance covered by the motor proteins before its detachment from the track has been defined as `processivity' and the influence of different motor intrinsic properties (amino-acid sequence \cite{pmid25109325,pmid18805095}, motor heads \cite{pmid10024239,pmid11086010}) and external factors (temperature \cite{nara_2006,pmid10860848}, track structure \cite{pmid19823565}) on processivity has been quantitatively studied by the measurement of motor speed and force generation. 


Despite the availability of extensive quantitative measurements, the rate of spontaneous motor protein detachment from the track has not been measured explicitly under diverse conditions where motor speed and force are known to be influenced. As a result, in different modelling studies the spontaneous detachment rate is considered to be independent of the motor speed \cite{pmid16486411,pmid17142285}. In the following, we demonstrate that the rate of spontaneous motor detachment from the track and its speed are correlated for Kinesin and Dynein. Further, using a one dimensional random walk model the influence of the spontaneous motor detachment from the track has also been presented. The effect of the motor speed and detachment rate relationship is studied in two conditions- time controlled setup and distance controlled setup. In these two conditions we obtain the distance travelled by the motor protein in a  given duration and the time taken by the motor protein to cover a pre-specified distance, respectively. 

\section{Materials and Methods}
\subsection{\label{subsec:motor_velocity_escape_propensity}Experimental observations: motor speed and detachment probability}
\begin{figure}[ht!]
\centering
\includegraphics[width=0.7\textwidth]{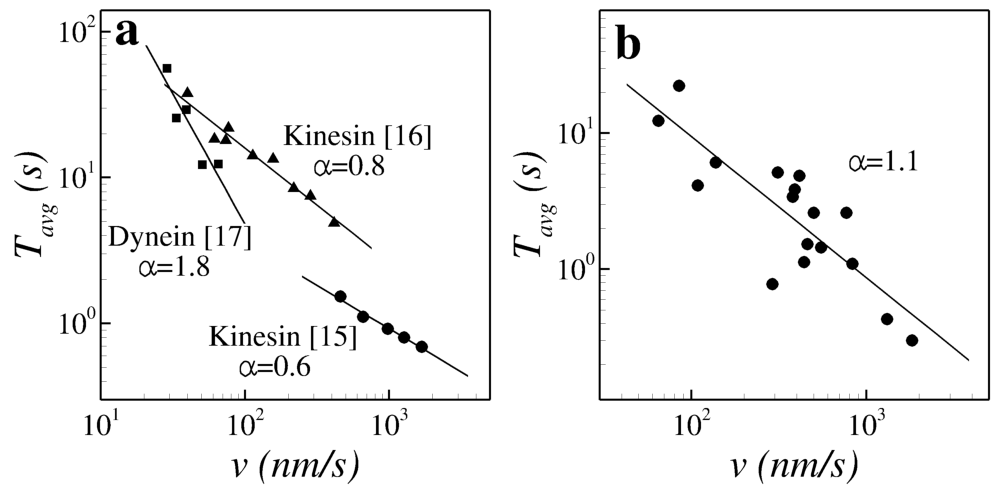}
\caption{(a) Relationship between experimentally measured motor speed, $v$, and average duration of its attachment to the track, $T_{avg}$ for Kinesin \cite{nara_2006, pmid18805095} and Dynein \cite{pmid25109325} motors. 
(b) The speed and the average attachment duration for different motor proteins to their respective tracks as per the experimental data available in literature. The lines in (a) and (b) show the power law fit $T_{avg} \sim v^{-\alpha}$. }
\label{fig:speed_detachment_expt}
\end{figure}

The distance based processivity, as defined previously, does not give any insight into the detachment probability, $p_e$, at any given time. To obtain the experimental estimates of the $p_e$, we looked into the literature where duration of the motor attachment to the protein tracks was measured along with its speed. In the study performed by Nara et al. \cite{nara_2006}, the kinesin was incubated with microtubules at different temperatures, whereas in the works by Cleary et al. \cite{pmid25109325} and Yildiz et al., \cite{pmid18805095}, the speeds and the attachment durations for dynein (with different mutations) and kinesin (with variable neck linker length) were measured, respectively (See Tables 
S1, S2 and S3 in supporting information for detailed experimental data).  When plotted on log-log scale, the average duration of the motor attachment to the track $T_{avg}$, and its speed reveal a monotonic relationship which can be approximated as a power law relation, $T_{avg} \sim v^{-\alpha}$ with $\alpha>0$ (Fig. \ref{fig:speed_detachment_expt}a). This correlation between the motor speed and rate of detachment was not highlighted in the respective studies \cite{nara_2006,pmid25109325,pmid18805095}. Furthermore, a compilation of $T_{avg}$ and the speeds of different motors in diverse experimental conditions also demonstrated similar relation with $\alpha=1.1$ (Fig. \ref{fig:speed_detachment_expt}b, see Table 
S4 in supporting information for experimental data). The positive values of $\alpha$ indicate a tradeoff between the motor speed and its duration of attachment to the track as an increase in the speed leads to faster detachment.  In the following we have utilized this speed-attachment duration relation to explore the transport of motor proteins.

\subsection{\label{subsec:distance_finite_time}Mathematical model}
\subsubsection{\label{subsubsec:assumptions}Drift-diffusion with detachment}
The detachment of motor protein from the track limits it from reaching its target destination, the end of the filament track within a finite time.
Therefore, an estimate of the distance covered by the motor protein in given duration (time controlled setup) and the time required for covering a prespecified distance (distance controlled setup) are crucial aspects for complete understanding of the motor driven cellular transportation.
%

%
In general, the motion of the motor proteins on track is constituted of forward (and in some cases backward too) steps of different sizes where each step is driven by ATP hydrolysis \cite{lodish}. Additionally, the motor protein can also spontaneously get detached from the track at any cycle with a detachment probability $p_e$. Considering the motion of the motor proteins on the tracks as a one-dimensional random walk, its position on the track at any given time can be characterized by the probability density $P(x,t)$ which follows the Fokker-Planck equation \cite{ding_ranga_2004}
\begin{equation} \label{eq:fokker-planck_equation}
\frac{\partial P(x,t)}{\partial t} = - v \frac{\partial P(x,t)}{\partial x} + D \frac{\partial^2 P(x,t)}{\partial x^2} - P(x,t) r_e
\end{equation}
where $v$ and $D$ are the motor speed and its diffusion constant on the track, respectively, whereas $r_e$ rate of motor detachment from the track and is related to the detachment probability $p_e$ as \cite{feldman}
\begin{equation}
r_e = -\log(1-p_e)/\tau,
\end{equation}
where $\tau$ is the time required for the hydrolysis of single ATP. It has to be noted that for small values of $p_e$, $r_e=p_e/\tau$. In order to understand the properties of the stochastic processes, the first passage time distribution is often used \cite{ding_ranga_2004}. In the context of the motor protein motion the first passage time distribution, $f(L,t)$, is defined as the time distribution for the motor protein crossing a target distance $L$. In other words, the probability that the motor protein will cover a distance $L$ between time $t$ and $t+dt$ is given by $f(L,t)dt$. Using the approach described in \cite{ding_ranga_2004} we can derive the first passage time distribution corresponding to the equation \ref{eq:fokker-planck_equation} as
\begin{equation} \label{eq:psi_continuum}
f(L,t) = \frac{L}{\sqrt{4\pi D t^3}} \exp\left(-r_e t-\cfrac{(L-vt)^2}{4Dt} \right)
\end{equation}
%
In the scenario where the motor does not detach from the track ($r_e=0$), the above expression reduces to the first passage time distribution for the Brownian particle with drift. In the following, we will use the first passage time distribution in the time and distance controlled setups.

\subsubsection{Time controlled setup: Distance travelled in fixed time}
A measure of a motor protein transport can be obtained in terms of the distance covered by it in a prespecified duration. The average distance covered in time $T$ can be written as 
\begin{equation} \label{eq:average_dist}
L_m(T) = \int \limits_{0}^{\infty} \int \limits_{0}^{T} L\left( \frac{\partial f}{\partial L}\right)dt dL.
\end{equation}
In general, the analytical evaluation of the above integral is not straight forward but it can be calculated for some special cases. For the purely ballistic motion, that is $D=0$, we can evaluate the above integral by substituting the first passage time distribution $f(L,t)=\delta\left(t-L/v \right)$ to obtain
\begin{equation} \label{eq:average_dist_D0}
L_m \left( T, D=0 \right) = \frac{v}{r_e} \left( 1 - e^{-r_e T}\right).
\end{equation}
On the other hand, for the purely diffusive motion ($v=0$), we obtain 
\begin{equation}\label{eq:average_dist_v0}
L_m \left( T, v=0 \right) = \sqrt{\frac{D}{r_e}} \text{erf} \left(\sqrt{r_e T}\right)
\end{equation}
where $\text{erf}(x)$ is the error function. From the relations \ref{eq:average_dist_D0} and \ref{eq:average_dist_v0}, it is apparent that $L_m$ attains finite values $L_s=v/r_e$ and $L_s=\sqrt{D/r_e}$, respectively, for $T \rightarrow \infty$ and is defined as processivity. 
\subsubsection{Distance controlled setup: Motor transport to fixed distance}
Similar to time controlled setup, the motor transport can also be analyzed in the distance controled setup where the rate of motor travel to a fixed distance is calculated. In this scenario, due to its detachment from the track the motor protein is not necessary to reach the pre-specified distance. Therefore, first we need to calculate the probability that the motor protein can reach the given distance. This probability is given by 
\begin{equation} \label{eq:reach_probability}
P_r(L) = \int \limits_{0}^{\infty} f(L,t)dt.
\end{equation}
In the case it can cover the distance ($P_r(L)>0$), the average time taken to achieve it can be written as
\begin{equation}\label{eq:average_time}
T_m(L) = \frac{1}{P_r(L)} \int \limits_{0}^{\infty} t f(L,t)dt.
\end{equation}
It has to be noted that for $r_e=0$, the above integral diverges but $r_e > 0$ results in finite $T_m$. Further, in this distance controlled setup the ratio $\phi(L) = P_r/T_m$ can be defined as the rate of motor transport to a fixed distance. The physical interpretation of $\phi(L)$ can be understood as the fraction of the motor proteins reaching a target distance per unit time. For instance, low transport rate ($P_r/T_m \rightarrow 0$) means that if a motor protein starts at some location of the track it does not reach the specified distance. This failure could either be due to the detachment of the motor protein from the track or due to its extremely slow motion along the track. On the other hand, $P_r/T_m$ takes high values only if $P_r$ is high and $T_m$ is low.

Similar to the time controlled setup, we can evaluate the probability and average time taken for reaching the target distance in the limiting cases. For ballistic motion, we obtain
\begin{align}
P_r \left( L, D=0 \right) &= e^{-r_e L/v}, \\
\text{and }T_m \left( L, D=0 \right) &= \frac{L}{v}.
\end{align}
For diffusive motion
\begin{align}
P_r \left( L, v=0 \right) &= \exp{\left( -\sqrt{\frac{L^2 r_e}{D}} \right)}, \label{eq:pr_v0}\\
\text{and }T_m \left( L, v=0 \right) &= \frac{L}{2\sqrt{ D r_e}}  \label{eq:tm_v0}.
\end{align}
Expectedly, for ballistic motion the average time $T_m$ is not dependent on the detachment probability. On the other hand for diffusive transport ($v=0$), we obtain $T_m \rightarrow 1/\sqrt{p_e}$ as $p_e \rightarrow 0$.
\subsubsection{\label{subsubsec:with_velocity_detachment}Incorporation of experimental observations}
As observed in the experiments (Fig. \ref{fig:speed_detachment_expt}), the motor speed and the rate of motor detachment from the track are not independent. For the power-law relations between the detachment probability ($p_e$, Fig. \ref{fig:speed_detachment_expt}A-B) we have 
\begin{equation} \label{eq:re_v_alpha}
\frac{1-e^{-r_e(v)\tau}}{1-e^{-r_e^0\tau}} = \left(\frac{v}{v_0}\right)^{\alpha}
\end{equation}
where $v_0$ and $r_e^0$ are the reference motor speed and the rate of detachment, respectively. We substitute the expression for $r_e$ from the above relation to obtain the average distance covered in finite time and the transport rate to a finite distance as a function of the exponent $\alpha$. 
\section{Results and Discussion}
\subsection{Parameter values}
In this paper we have taken the transport of kinesin motor on microtubule as an example to understand the effects of the speed dependent detachment rate. The values of the model parameters for kinesin motor are shown in Table \ref{tab:paramvals}.
\begin{table}[h!]
  \centering
  \begin{tabular}{p{2cm} p{2cm} p{2cm}}
    \textbf{Parameter} & \textbf{Value} & \textbf{Reference}\\
    \hline
    $v_0$ & $640~nm/s$ & \cite{stryer_biochem}\\
		\hline
    $p_e^0$, $r_e^0$ & $0.01$, $0.35/s$ & \cite{nara_2006}\\
		\hline
    $\tau$ & $1/80~s$ & \cite{stryer_biochem}\\
		\hline
    $D$ & $2500~nm^2/s$ & \cite{Lu2009}
  \end{tabular}
	\caption{Values of the model parameters for kinesin motor used in the current study} \label{tab:paramvals}
\end{table}
The similar analysis can also be performed for dynein and myosin motor proteins.

\subsection{\label{subsec:motor_detachment}Influence of motor detachment from the track}
\begin{figure}[ht!]
\centering
\includegraphics[width=\textwidth]{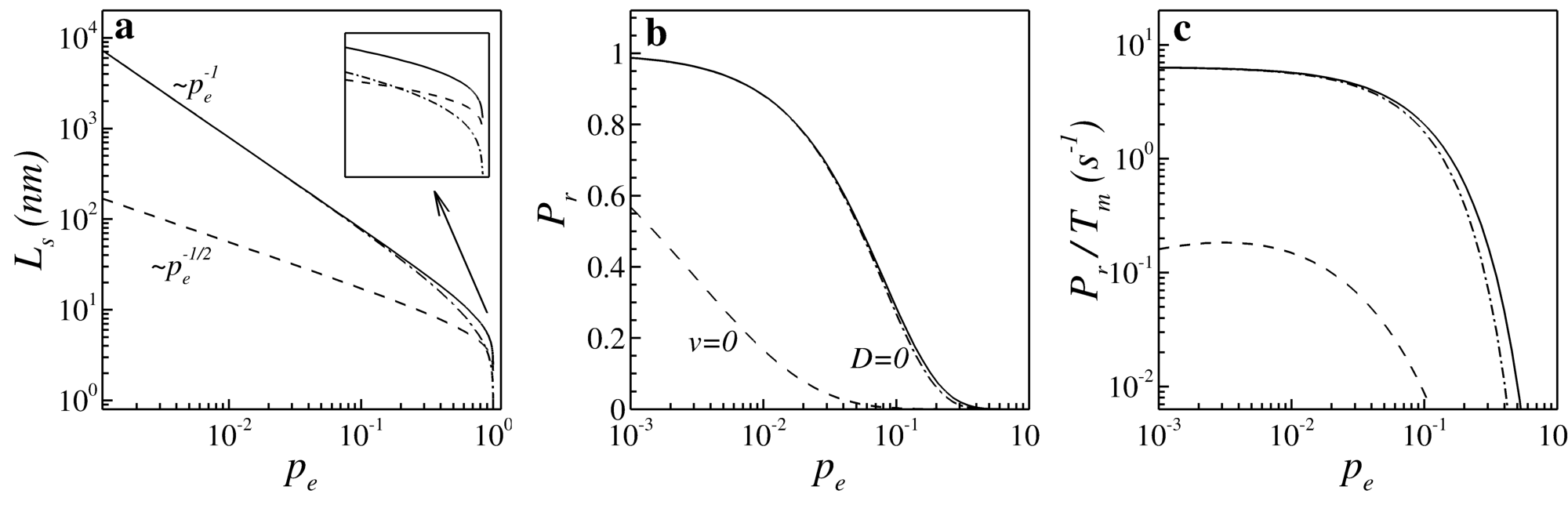}
\caption{Effect of kinesin detachment from microtubule on its transport for fixed speed and diffusivity. (a) Dependence of kinesin processivity $L_s$ under diffusive transport ($v=0$ and $D=2500nm^2/s$, dashed line), ballistic transport ($v=640nm/s$ and $D=0$, dash-dotted line), and mixed transport ($v=640nm/s$, $D=2500nm^2/s$, solid line) on detachment probability $p_e$. Inset shows the enlarged version of the same plot for large $p_e$. (b) Probability, $P_r$, of covering a distance of $100nm$ by kinesin on a microtubule track before detachment under diffusive (dashed), ballistic (dash-dotted) and mixed (solid) modes of transport. (c) Rate of kinesin transport on microtubule to a distance of $100nm$ for aforementioned three modes of transport.}
\label{fig:detach_LP}
\end{figure}

The detachment of the motor protein from the track limits the distance it can cover in finite time and the rate of its transport to a specific distance. In order to study the effect of the change in velocity with the detachment probability we first calculated the processivity of the motor protein when its velcoity does not change with $p_e$. Thus calculated processivity of a kinesin motor on microtubule track for fixed velocity and diffusivity is shown in Fig. \ref{fig:detach_LP}a. If the motor velocity remains fixed, for small values of $p_e$ its processivity decreases as $1/p_e$ and $1/\sqrt{p_e}$ for velocity and diffusivity dominant transports, respectively. This decrease in processivity, however, is contrary to the experimental observations where it shows an increase with $p_e$ \cite{nara_2006,pmid18805095}. This shows that the assumption of fixed rate of detachment \cite{pmid16486411,pmid17142285} independent of motor speed may not be correct. 

In the distance controlled setup, for fixed velocity and diffusivity the probability of motor protein reaching a target distance decreases with $p_e$ (Fig. \ref{fig:detach_LP}b) in both of the diffusion and drift dominated modes of transport. The transport rate to the target distance $\phi(L)$, however, shows different behaviors in diffusion and drift dominated motions (Fig. \ref{fig:detach_LP}c). For purely drift driven motor protein transport, the transport rate decreases with $p_e$. But in the diffusion driven transport, the transport rate follows a non-monotonic relationship with $p_e$. This is because for large $p_e$, the probability of reaching the target distance is negligible whereas for very small $p_e$ the the time taken for complete transport is very large (see relations \ref{eq:pr_v0} and \ref{eq:tm_v0}). The disagreement between the experimental observations of increased processivity with the detachment rate and model predictions with constant motor speed suggests that the dependence of motor speed on the detachment rate has significant effect on its transport properties. In the following we will study the effect of motor speed-detachment rate relationships on its transport under time and distance controlled conditions.

\subsection{Effect of speed dependent detachment rate}
\subsubsection{\label{subsubsec:timecontrolled}Motor transport in fixed time}
\begin{figure}[ht!]
\centering
\includegraphics[width=0.95\textwidth]{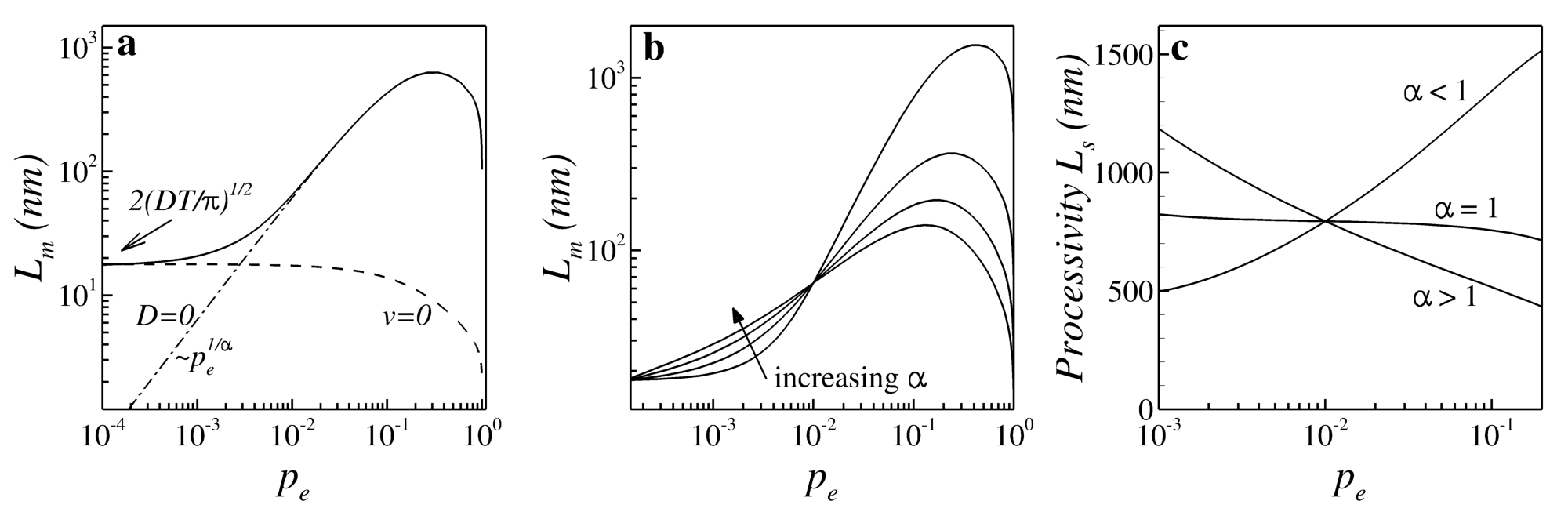}
\caption{Distance travelled by kinesin on microtubule in $100 \mu s$ with velocity dependent detachment rate. (a) Distance covered under diffusive (dashed), ballistic (dash-dotted) and mixed (solid line) modes of transport with $\alpha=1$. (b) Distance covered by kinesin for $\alpha=0.8,1.2,1.6,2.0$. (c) Dependence of motor protein processivity on $p_e$ for different $\alpha$.}
\label{fig:timeControl}
\end{figure}

The experimental data showed (Fig. \ref{fig:speed_detachment_expt}) that the motor speed and its detachment rate can be characterized as power law with exponent $\alpha$. With this power-law relationship, the distance covered by the motor protein in finite time (Fig. \ref{fig:timeControl}a) shows a non-monotonic dependence on $p_e$ which is unlike monotonically decreasing dependence when the motor speed does not change (Fig. \ref{fig:detach_LP}a). For very small values of $p_e$ the motor speed goes to zero and the distance covered in finite time takes a constant value ($L_m(T) = 2 \sqrt{DT/\pi}$) which is determined by the diffusivity of the motor protein alone. On the other hand, for drift dominated scenario ($D=0$) the distance covered in finite time increases as $p_e^{1/\alpha}$ for small values of $p_e$. For large $p_e$, the motor protein get detached from the track very fast and does not traverse any distance. Fig. \ref{fig:timeControl}b shows the distance covered by the kinesin on microtubule in finite time for different $\alpha$. Apart from the maximum value attained by $L_m$, the nature of the dependence of $L_m$ on detachment probability does not change with the power law exponent $\alpha$. As $T \rightarrow \infty$ the distance covered by the motor on track converges to its processivity, $L_s$ and it is shown in Fig. \ref{fig:timeControl}c. As opposed to the distance covered in finite time, the qualitative nature of processivity depends significantly on $\alpha$. It can be seen that for the motor proteins with $\alpha < 1$, such as kinesin, the processivity increases with the detachment probability, whereas for motor proteins like dynein, for which $\alpha > 1$, it decreases. This dependence of processivity on $\alpha$ is evident from the substitution of relation \ref{eq:re_v_alpha} in \ref{eq:average_dist_D0} for drift dominated transport, which gives for small $p_e$
\begin{equation}
L_s = \frac{v_0\tau}{\left(p_e^0\right)^{1/\alpha}} p_e^{(1-\alpha)/\alpha}. 
\end{equation}
This can be understood by comparing the relative changes in the velocity and detachment probability. For $\alpha < 1$, an increase in detachment probability leads to much larger increase in the motor velocity and hence increased processivity and vice a versa. Therefore, the power law exponent $\alpha$ describes the nature of the processivity of the motor proteins.

\subsubsection{\label{subsubsec:distancecontrolled}Motor transport to fixed distance}
\begin{figure}[ht!]
\centering
\includegraphics[width=0.7\textwidth]{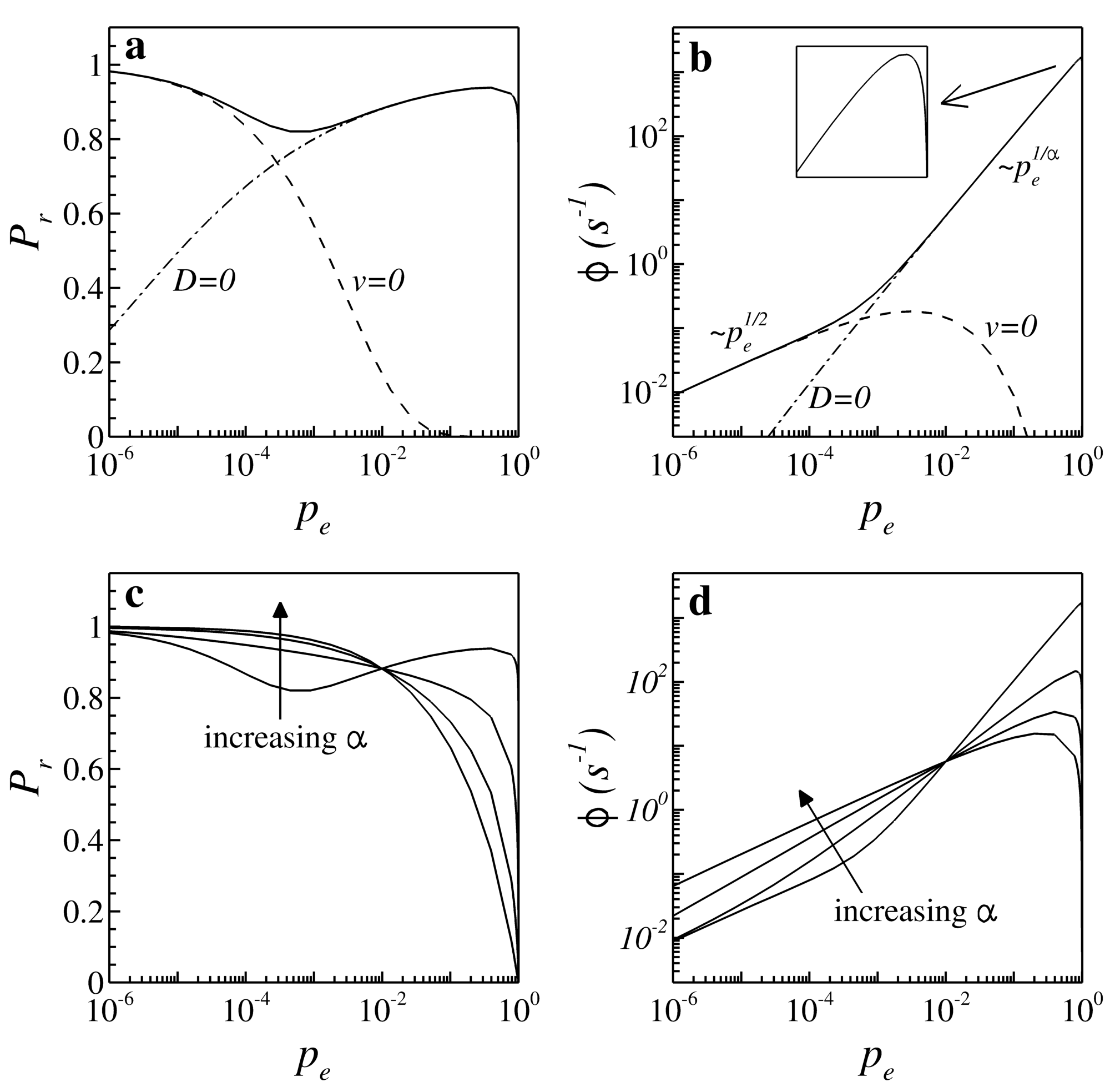}
\caption{Transport of kinesin to a fixed distance with velocity dependent detachment rate. (a) Probability $P_r$ of reaching and (b) rate of transport of kinesin motor to distance of $100\mu m$ before detachment under three modes of transport. (c) The probability $P_r$ and (d) the transport rate to the target distance of $100\mu m$ for $\alpha=0.8,1.2,1.6,2.0$.}
\label{fig:distanceControl}
\end{figure}

Similar to the track bound transport of motor protein in finite time we can also study the effect of the $v-p_e$ relations (Fig. \ref{fig:speed_detachment_expt} and equation \ref{eq:re_v_alpha}) on its transport to a fixed distance. The probability of reaching a target distance by the kinesin motor on microtubule is shown in Fig. \ref{fig:distanceControl}a which shows that for drift dominated scenario ($D=0$) the probability $P_r$ attains a maxima. This is opposed to the scenario when the motor speed does not change with the detachment probability (Fig. \ref{fig:detach_LP}b). On the other hand, for the rate of motor transport $\phi(L)$ both of the drift and diffusion dominated regimes attain local maxima (Fig. \ref{fig:distanceControl}b). Further, for small values of $p_e$ the rate of transport $\phi$ increases as $\sqrt{p_e}$ and $p_e^{1/\alpha}$ in the drift and diffusion dominated motions of the motor protein on the track, respectively. Figs. \ref{fig:distanceControl}c,d show the dependence of $P_r$ and $\phi$ on $p_e$ for difference values of $\alpha$. It has to be noted that the maxima attained by the $P_r$ vanishes for $\alpha > 1$. However, the rate of transport $\phi$ shows a maximal transport for all values of $\alpha$. This demonstrate that for track bound transport of motor proteins to a fixed distance there exists an optimal velocity which maximizes its rate of transport to the target location. 

\subsubsection{\label{subsubsec:optimumVelocity}Optimum motor velocity}
As shown in Fig. \ref{fig:timeControl}b and Fig. \ref{fig:distanceControl}d the track bound transport of the motor proteins has non-monotonic dependence on the motor speed (or detachment rate). Therefore, faster motor speed on a track does not necessarily mean its faster transport and it is possible to obtain the motor velocity corresponding to the farthest transport in fixed time and fastest transport to a pre-specified distance.

For kinesin, its velocity in physiological conditions $v=640nm/s$ turns out to be optimal for its farthest transport in time $T \sim 10s$. For the cellular functions, this duration is too large. 
On the other hand, the same physiological velocity of kinesin is optimal for the fastest transport to a distance of $L > 5 \mu m$. This distance is roughly of the same order of magnitude as that of the size of a single cell. 
This demonstrates that kinesin motor is more likely to move at speeds to maximize the transport rate to a given distance instead of maximizing the distance in given time. This observation is reasonable as in physiological conditions the transport of cellular cargo is to fixed distances as the cell size remains more or less unchanged. 

It has to be noted that in physiological conditions motor protein does not constantly move on a single track. For instance, in case of kinesin transport on microtubule tracks it keeps hopping among different microtubule filaments \cite{Ross2008} and also undergoes diffusive motion in cytosol. Furthermore, consumption of ATP is another factor governing the motor transport in biological systems. From the economy viewpoint, the optimal motor velocity can also be determined by the one which requires minimal ATP consumption. Therefore the estimation of optimal speed of the motor protein has to take these factors into account which is not in the scope of present study. 


\subsection{Biological implications and future directions}
In this paper, using existing experimental data, we have demonstrated a novel monotonic relationship between the speed of the motor protein and the duration of its attachment to the track as a tradeoff between the two. Although an insight into the exact nature of this relationship will require a detailed experimental and modelling endeavour, a physical understanding of its monotonic nature for bipedal motors can be obtained by the following argument. The kinesin and other bipedal motors move on the tracks by the sequential attachment and detachment of the two heads \cite{pmid9335494}. The speed and the detachment probability depends on the duration of attachment of each motor head in each cycle. In case of high (low) motor speed, its both heads undergo fast (slow) attachment-detachment transitions which also leads to high (low) detachment probablity. For the range of the motor speeds presented here, a power-approximation leads to an easy prediction that increase in the speeds of the kinesin ($\alpha < 1$) and dynein ($\alpha > 1$) motors would result in higher and lower processivity, respectively. For kinesin, it has already been observed experimentally by varying motor speed by temperature changes \cite{pmid10860848}. However, the same is yet to be experimentally verified for dynein. 

Further, the observation of the tradeoff is based on different experimental studies performed under diverse experimental conditions. Therefore, a systematic experimental study is warranted to test the conditions under which the speed-detachment tradeoff holds. To achieve this the model predictions presented in this paper can be utilized. 

It has to be noted that although the exponent $\alpha$ is the characteristic of the motor protein, it also depends on its microenvironment. This can be seen from the experimental data shown in Figs. \ref{fig:speed_detachment_expt}a where the average duration of attachment of the kinesin motor to the microtubule track is obtained by two groups (\cite{pmid25109325} and \cite{nara_2006}) differs significantly if not by an order of magnitude. Similarly, in Fig. \ref{fig:speed_detachment_expt}B also we observe multiple values of the attachment duration and the speed of the kinesin motors. The reason for this discrepancy is the differing experimental conditions used in each setup and hence this aspect has to be kept in consideration during classification of motor proteins based on $\alpha$. Similar speed-accuracy tradeoffs have also been observed in the context of DNA replication \cite{pmid25554788}, road-safety \cite{nilsson2004thesis}, human movements \cite{pmid10096999,pmid18217850}, decision making \cite{wickelgren1976,pmid24966810} and perceptual learning\cite{pmid21958757}. Although, the underlying mechanisms of the speed-accuracy tradeoffs might be different in different systems, its influence on the overall system dynamics would still hold.

Furthermore, it is known that motor proteins often undergo collective motions on the tracks. Therefore, it would be interesting to see the effect of the speed-detachment tradeoff on the collective motion of the motor proteins. Additionally, the model can also be extended to study the effect of speed-detachment tradeoff on the motor transport in more realistic scenario where the motor protein can also reattach to the track. 

\section{Conclusion}
This work utilizes the available experimental observation to propose a novel relationship between the speed of motor proteins and the rate of their detachment from the track. This tradeoff is further utilized to study the characteristics of the motor transport along the track in fixed time and to a fixed distance. It is shown that faster motor speed does not necessarily mean faster or farthest transport. Therefore, optimal speed can be estimated for the transport of motor proteins along the cytoskeletal filaments.

\end{document}